\documentclass[aps, twocolumn, superscriptaddress]{revtex4} 

\usepackage[switch]{lineno}
\usepackage{amsmath}
\usepackage{empheq}
\usepackage[parfill]{parskip}
\usepackage{array}
\usepackage{booktabs}
\usepackage{amsfonts}
\usepackage{dsfont}
\usepackage{graphicx}
\usepackage{amsthm}
\usepackage{hyperref}
\usepackage{multirow}
\usepackage{tcolorbox} 
\usepackage{scrextend}

\begin{document}

\setlength{\parindent}{0.5cm}

\title{Effects of coupling range on the dynamics of swarmalators}
\author{Gourab Kumar Sar}
\email{mr.gksar@gmail.com}
\affiliation{Physics and Applied Mathematics Unit, Indian Statistical Institute, 203 B. T. Road, Kolkata 700108, India}
\
\author{Kevin O'Keeffe}
\email{kevin.p.okeeffe@gmail.com}
\affiliation{Starling Research Institute, Seattle, USA} 
\author{Dibakar Ghosh}
\email{dibakar@isical.ac.in}
\affiliation{Physics and Applied Mathematics Unit, Indian Statistical Institute, 203 B. T. Road, Kolkata 700108, India}

\begin{abstract}

We study a variant of the one-dimensional swarmalator model where the units' interactions have a controllable length scale or range. We tune the model from the long-range regime, which is well studied, into the short-range regime, which is understudied, and find diverse collective states: sync dots, where the swarmalators arrange themselves into $k>1$ delta points of perfect synchrony, $q$-waves, where the swarmalators form spatiotemporal waves with winding number $q>1$, and an active state where unsteady oscillations are found. We present the phase diagram and derive most of the threshold boundaries analytically. These states may be observable in real-world swarmalator systems with low-range coupling such as biological microswimmers or active colloids. \\

\noindent
DOI: XXXXXXX
\end{abstract}


\maketitle

\section{Introduction}
Swarmalators are oscillators capable of moving around in space \cite{o2017oscillators}. They model many systems that mix swarming with synchronization, such as biological microswimmers \cite{adorjani2024motility}, magnetic domain walls \cite{hrabec2018velocity}, Japanese tree frogs \cite{aihara2014spatio}, migratory cells \cite{riedl2023synchronization}, active spheres \cite{riedl2023synchronization,riedl2024entropy}, active spinners \cite{sungar2024synchronization}, and robotic swarms \cite{barcis2019robots,barcis2020sandsbots}. 

The first studies of swarmalators considered the simplest case of uniform coupling and found synchronous disks and vortices \cite{o2017oscillators}. Later studies sought a deeper understanding of this foundational model \cite{ha2021mean,ha2019emergent,gong2024approximating,degond2022topological} while other studies have generalized the model. Some swapped the uniform coupling with more realistic coupling schemes, such as those with delays \cite{blum2024swarmalators,sar2022swarmalators,lizarraga2023synchronization}, stochastic failure rates \cite{schilcher2021swarmalators}, mixed sign interactions \cite{hao2023attractive,hong2021coupling,o2022swarmalators} or multiple Fourier harmonics \cite{smith2024swarmalators}, or non-pairwise interactions \cite{anwar2024collective}. Others added new features into the model, such as random pinning~\cite{sar2023pinning,sar2023swarmalators,sar2024solvable}, external forcing \cite{lizarraga2020synchronization,anwar2024forced}, environmental noise \cite{hong2023swarmalators}, and geometric confinement \cite{degond2023topological}. The effect of non-Kuramoto type phase dynamics has also been considered in a recent study~\cite{ghosh2024amplitude}. Researchers have also been keen on studying these systems with multiplex~\cite{kongni2023phase}, modular network structures~\cite{ghosh2023antiphase} and other effects.~\cite{sar2022dynamics,o2019review}.

Yet in all the studies above, the swarmalators' interactions were long-range (usually accompanied by a short-range hard shell repulsion term). Swarmalators with short-range interactions are less studied, though they occur in many real-world systems. Robotic drones and rovers driven with swarmalator control, for instance, have sensors with fixed reach, and active spheres \cite{riedl2024entropy} only communicate when they physically touch.

Short-range coupling in swarmalator systems were first studied by Lee \cite{lee2021collective}, Bettstetter \cite{schilcher2021swarmalators}, and Ceron \cite{ceron2022diverse}. They added finite-cutoff range $\sigma$ in the original two-dimensional (2d) swarmalator model and found several new emergent states. Even so, their studies were purely numerical. The 2d model, while minimal, is still too difficult to analyze (see Introduction in \cite{o2023solvable} for a discussion about why) leaving a theoretical understanding of swarmalators with short-range coupling lacking.

This paper sets out to address this research gap. Our strategy is to retreat one spatial dimension and attach a tunable sensing range to the 1d swarmalator model \cite{o2022collective,yoon2022sync}. The one-dimensional (1d) swarmalator model restricts the swarmalators' motion to a 1d periodic domain which makes it one of the few models that is tractable \cite{o2024stability} (it turns it into Kuramoto-like model which means we can use techniques from sync studies to attack it). It has facilitated several exact analyses of swarmalators \cite{o2022collective, yoon2022sync, o2024stability,  o2024global, hao2023attractive, o2022swarmalators,sar2023pinning}. Here it allows us to analytically determine the critical coupling range at which various collective states arise and destabilize. These are the first analytic results about coupling range on swarmalators and in that sense help advance the field.

\section{Model} \label{model}
\begin{figure}[t]
	\centering
	\includegraphics[width=\columnwidth]{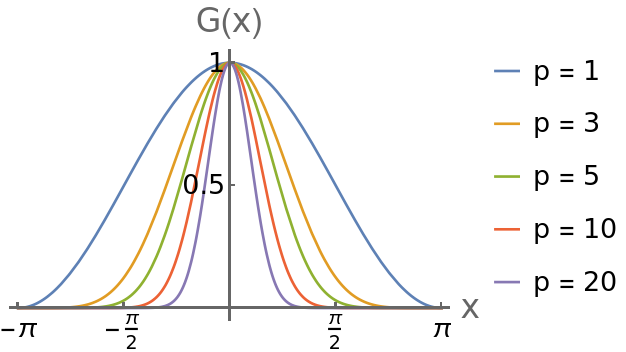}
	\caption{The function $G(x) = (\frac{1 + \cos x }{2})^p$ is plotted for different values of $p$. The larger the value of $p$, the smaller the region of $x$ where its value is nonzero. It effectively works as a local coupling for large $p$ values.}
	\label{pulse}
\end{figure}

The 1d swarmalator model is
\begin{align}
    \dot{x_i} &=  v + \frac{J}{N} \sum_j^N \sin(x_j - x_i) \cos(\theta_j - \theta_i) G(x_j - x_i), \label{xdot}\\
    \dot{\theta_i} &= \omega + \frac{K}{N}  \sum_j^N \sin(\theta_j - \theta_i ) \cos(x_j - x_i ) G(x_j - x_i) , \label{thetadot}
\end{align}
\noindent
where $(x_i, \theta_i) \in (\mathbb{S}^1, \mathbb{S}^1)$ are the position and phase of the $i$-th swarmalator, $J$, $K$ are the associated couplings, and $v$, $\omega$ the associated frequencies. By going to a suitable frame, we can set $\omega=\nu=0$ to zero without loss of generality. This model has been studied before \cite{o2022collective}. The novelty here is the kernel $G(x)$ contains the coupling range (for simplicity we assume the same $G$ for both the $\dot{x}$ and $\dot{\theta}$ equations) which has form
\begin{align}
G(x) = \Big( \frac{1 + \cos x }{2} \Big)^p.
\end{align}
Figure~\ref{pulse} shows this is a pulse with controllable width $p$ where larger $p$ corresponds to narrower widths. We choose such a $G(x)$ because it is nice to work with analytically. For a given range $p$, it has a finite number of Fourier modes; when you project onto the Fourier basis the calculation comes out clean. Other kernels such as the box function do not have this convenient feature.

To recap, we have a model with three parameters $(J,K,p)$, where the sensing range / pulse width $p$ is new. We recover the baseline model in the limit $p=0$.

To get a feel for how our model behaved, we ran simulations for different $p$ and found several new collective states. We first present and analyze these states for fixed $p=2$, and variable $(J,K)$. This is a convenient way to present our analysis because then the functional form of the model is fixed ($G(x)$ contains a fixed number of harmonics). Moreover, the physics effectively doesn't change for an increased $p$ (better said, it changes in a predictable way). So the $p=2$ case captures the essential phenomenology. After that, we generalize our analysis to variable $p$.
\section{Results for pulse width $p=2$}
The following generalized Kuramoto order parameters~\cite{kuramoto1975self} will be useful
\begin{align}
	Y_n &= Q_n e^{i \Psi_n} = \frac{1}{N} \sum_j e^{i n x_j} ,\nonumber\\
	Z_n &= R_n e^{i \Theta_n} = \frac{1}{N} \sum_j e^{i n \theta_j} ,
\end{align}
for $n \in \mathbb{Z}^+$. $R_1$ captures the amount of phase coherence and is the most important of them all in our model. We also use the rainbow order parameters~\cite{o2022collective}
\begin{equation}
   W_{n \pm} = S_{n\pm} e^{i \Phi_{n \pm}}=  \frac{1}{N} \sum_j e^{i (n x_j \pm \theta_j )} , \label{rainbow}
\end{equation}
that measure the correlation between the swarmalators' spatial positions and phases. 

When $p=2$ our model is the following pair of governing equations
\begin{align}
    \dot{x_i} &=  \frac{J}{N} \sum_j^N \sin(x_j - x_i) \cos(\theta_j - \theta_i) \Big( \frac{1 + \cos (x_j - x_i }{2} \Big)^2, \label{x2dot}\\
    \dot{\theta_i} &= \frac{K}{N}  \sum_j^N \sin(\theta_j - \theta_i ) \cos(x_j - x_i ) \Big( \frac{1 + \cos (x_j - x_i }{2} \Big)^2 . \label{theta2dot}
\end{align}
In terms of the order parameters the model above can be rewritten as
\begin{multline}
    \dot{x_i} = \frac{J}{32}\Big(5 S_{1-} \sin \left[\Phi_{1-} - (x_i-\theta_i )\right]  \\+4 S_{2-} \sin \left[\Phi_{2-} - (2 x_i-\theta_i )\right] +S_{3-}\sin\left[\Phi_{3-} - (3 x_i-\theta_i )\right]  \\+5 S_{1+} \sin\left[\Phi_{1+} - (x_i + \theta_i)\right] +4 S_{2+}\sin\left[\Phi_{2+} - (2 x_i + \theta_i)\right]  \\ + S_{3+} \sin\left[\Phi_{3+} - (3 x_i + \theta_i)\right]\Big), \label{eq11}
\end{multline}
\begin{multline}
    \dot{\theta_i} = \frac{K}{32} \Big( -7 S_{1-} \sin\left[\Phi_{1-} - (x_i-\theta_i )\right] \\ -4 S_{2-} \sin\left[\Phi_{2-} - (2 x_i-\theta_i )\right] - S_{3-} \sin\left[\Phi_{3-} - (3 x_i - \theta_i )\right]  \\+7 S_{1+} \sin\left[\Phi_{1+} - (x_i + \theta_i)\right] +4 S_{2+} \sin\left[\Phi_{1+} - (2 x_i + \theta_i)\right] \\ + S_{3+} \sin\left[\Phi_{1+} - (3 x_i + \theta_i)\right]+8 R_1 \sin\left[\Psi_1 - \theta_i \right]\Big). \label{eq12}
\end{multline}

Numerical simulation of this model revealed the following collective states
\begin{figure}[t]
	\centering
	\includegraphics[width=1\columnwidth]{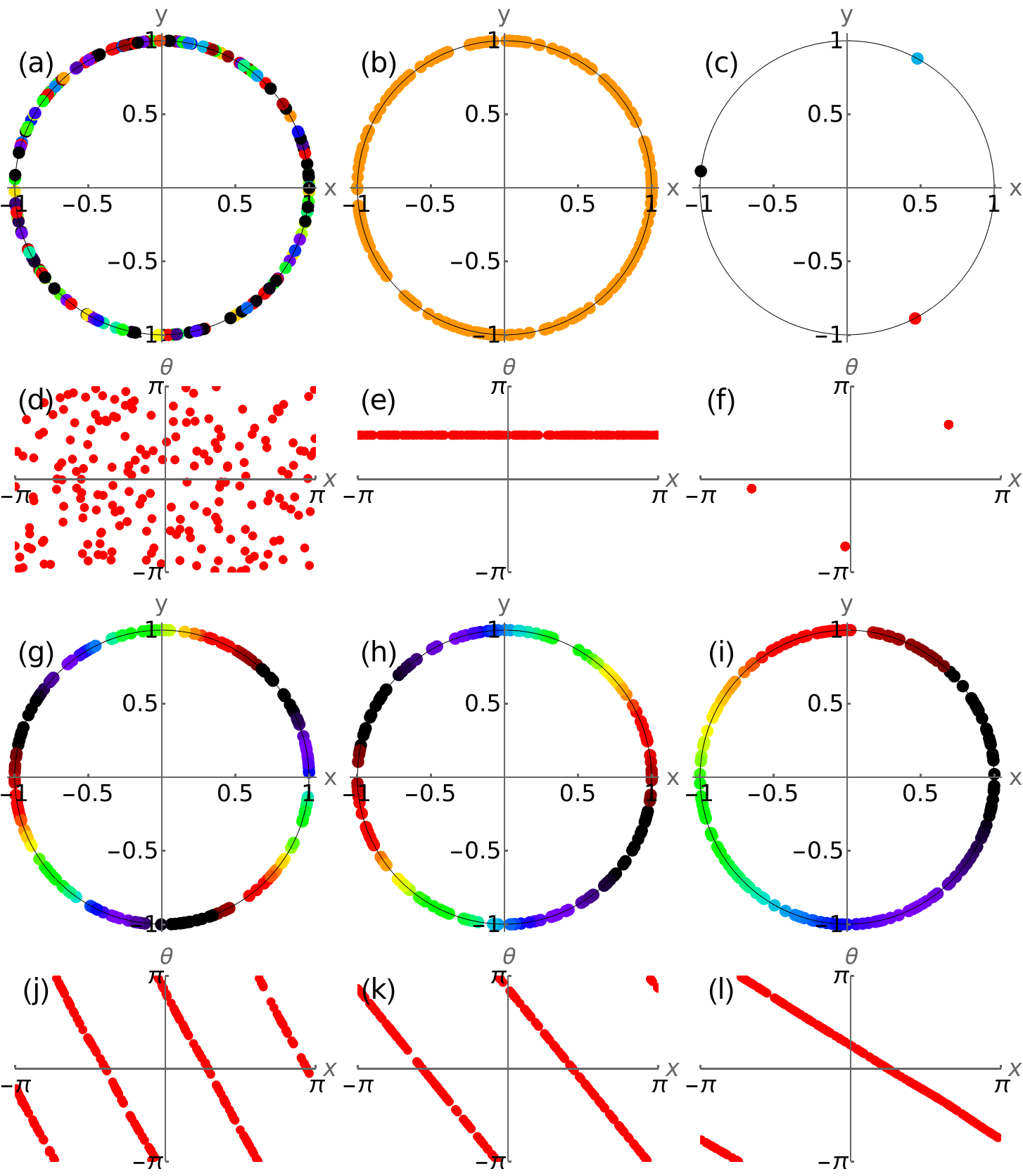}
	\caption{Collective states of the pulse model with $p=2$. (a,d) Async state $(J,K)=(2.5,-10)$. (b,e) Sync wave $(J,K) = (-2.5,1)$. (c,f) Sync 3-dots $(J,K)=(2.5,1)$. (g,j) 3-wave $(J,K)=(2.5,-6)$. (h,k) 2-wave $(J,K)=(2.5,-3)$. (i,l) 1-wave $(J,K)=(2.5,-0.1)$. At $t=0$, positions and phases are chosen randomly from $[0,2\pi]$ for all the collective states. $(t,dt,N) = (200,0.1,200)$.}
	\label{states-pulse}
\end{figure}

{\bf Async}. Swarmalators are fully desynchronized in both positions and phases. The correlation between the phases and positions is also lacking. Order parameter values lie near zero, i.e., $S_{1+}=0, S_{1-} =0$. See the plots in Figs.~\ref{states-pulse}(a,d). This is found in the baseline model.

{\bf Sync wave}. The swarmalators spread out uniformly over the ring in a wave with their phases fully synchronized. See Figs.~\ref{states-pulse}(b,e). This state is novel.

{\bf Sync dots}. Now the swarmalators settle into  $q>1$ zero-dimensional fully synchronized `dots' spaced equally apart: $x_i, \theta_i \in \{\frac{2\pi j}{q} + C : j=1,2,\dots,q\}$~\footnote{the constant depends on the initial conditions and the same is true for all other such constants that appear in the fixed points}. Figure~\ref{states-pulse}(c) shows the $q=3$ dots. The $q=1,2$ dots occurs in the base model (recovered in the limit $p=0$). Dots with $q>2$ are new. 

{\bf $q$-waves}. The swarmalators spread out in a wave over the ring again, but this time their phases are correlated with their positions $\theta_i = \pm q x_{i} + C$. The $q=1$ wave occurs in the base model. Winding number $q>1$ are new. The $q_{max}$ observed depends on the value of $p$ as we discuss later. Figures~\ref{states-pulse}(g)-(i) depict a few of these waves.

\begin{figure}[htp]
	\centering
	\includegraphics[width=\columnwidth]{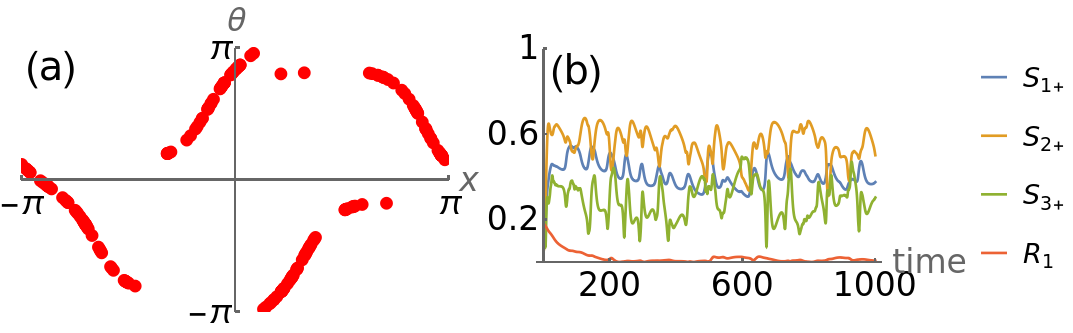}
	\caption{Active state near the boundary between the 1-wave and sync dots. (a) Swarmalators are shown in the $x$-$\theta$ plane. (b) The time evolution of the order parameters $S_{1+}$, $S_{2+}$, $S_{3+}$, and $R_1$ dictates the activity of the state. Coupling parameters are $(J,K=2.5,-0.1)$. Simulation parameters: $(t,dt,N)=(1000,0.1,200)$.}
	\label{active-state}
\end{figure}

{\bf Active state}. Apart from these static states, we find the emergence of an active state where the positions and phases of the swarmalators keep evolving over time and the oscillations never die. This state occurs near the boundary of the 1-wave and the sync dots. The scatter plot of this state in the $x$-$\theta$ plane is delineated in Fig.~\ref{active-state}(a). The activity is represented through the order parameters $S_{n+}$ and $R_1$ in Fig.~\ref{active-state}(b).


\begin{table*}
	\caption{\label{table1}Description of the collective states through the values of the order parameters. The first set of order parameters $S_{n\pm}$ indicates the correlation of type $\theta = \pm nx +C$, whereas, the second set of order parameters $Q_n,R_n$ is the measure of spatial and phase synchronization, respectively.}
	\begin{ruledtabular}
		\begin{tabular}{ccccccccccccc}
			&\multicolumn{6}{c}{$S_{n\pm}$\footnote{For convenience, we take $S_{n+}=\max\{S_{n+},S_{n-}\}$, and $S_{n-}=\min\{S_{n+},S_{n-}\}$ as otherwise they vary depending on the initial conditions. }}&\multicolumn{6}{c}{$Q_n,R_n$}\\
			State&$S_{1+}$&$S_{1-}$&$S_{2+}$&$S_{2-}$&$S_{3+}$&$S_{3-}$&$Q_1$&$Q_2$&$Q_3$&$R_1$&$R_2$&$R_3$\\ \hline
			Async &0&0&0&0&0&0&0&0&0&0&0&0 \\
			$1$-wave &1&0&0&0&0&0&0&0&0&0&0&0 \\
			$2$-wave &0&0&1&0&0&0&0&0&0&0&0&0 \\
			$3$-wave &0&0&0&0&1&0&0&0&0&0&0&0 \\
			Sync wave &0&0&0&0&0&0&0&0&0&1&1&1 \\
			Sync 1-dot &1&1&1&1&1&1&1&1&1&1&1&1 \\
			Sync 2-dots &1&1&0&0&1&1&0&1&0&0&1&0 \\
			Sync 3-dots &1&$M_1$\footnote[2]{$0<M_3<M_2<M_1\ll 1$. The exact values are not much relevant as we only use the fact that they are much less than 1.}&1&$M_2$\footnotemark[2]&$M_1$&$M_3$\footnotemark[2]&$M_2$&$M_1$&1&$M_2$&$M_1$&1 \\
		\end{tabular}
	\end{ruledtabular}
\end{table*}

\begin{figure}[htp]
	\centering
	\includegraphics[width=\columnwidth]{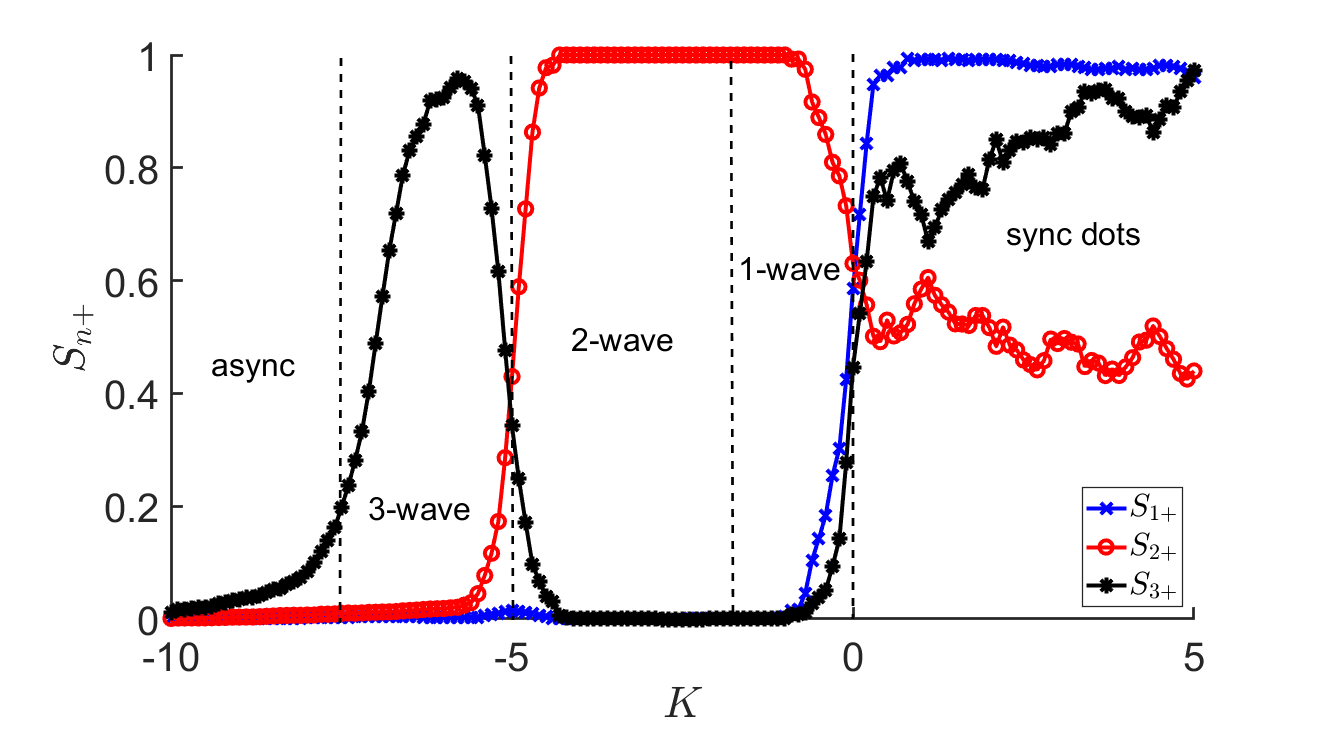}
	\caption{We fix $J=2.5$. Order parameters as functions of $K$. At $t=0$, positions are phases are chosen randomly from $[0,2\pi]$. $(t,dt,N) = (200,0.2,200)$. We take $10$ realizations over which the order parameters are averaged.}
	\label{op-pulse}
\end{figure}

Figure~\ref{op-pulse} illustrates the transitions between some of the states by plotting the dependence of $S_{n\pm}$ versus $K$. We take $n=1,2,3$ and use the convention that $S_{n+} = \max \{S_{n+},S_{n-}\}$. Starting at $K=-10$, first we see that $S_{1+}$ (blue crosses), $S_{2+}$ (red circles), and $S_{3+}$ (black stars) all lie close to zero, that indicate the async state. Around $K=-7.5$, $S_{3+}$ bifurcates from zero and we confirm the appearance of $3$-wave. Next, as $K$ approaches the value $-5$, $S_{3+}$ falls to zero and $S_{2+}$ bifurcates from zero. We arrive at the $2$-wave state where the solutions follow $\theta_i = \pm 2 x_i +C$ for some constant $C$. Further increasing the value of $K$, we observe that $S_{2+}$ starts to decrease from 1 and $S_{1+}$, $S_{3+}$ start to increase from zero just before $K=0$. In this region we report the bistability between $1$-wave and $2$-wave. We discuss the bistability in details in Sec.~\ref{bistable}. Finally, after $K=0$, we arrive at the sync dots where the sync 1-dot, sync 2-dots, and sync 3-dots are all probable. $S_{1+}$ always stays at $1$ while $S_{2+}$ and $S_{3+}$ acquire some nonzero values. The black dotted vertical lines in Fig.~\ref{op-pulse} stand for the analytically calculated values (see Sec.~\ref{analytical}) where these transitions occur.

As we can already realize that we have to deal with quite a few order parameters to characterize these states, in Table~\ref{table1} we write down their values in the different collective states.

\subsection{Stability analysis of the collective states} \label{analytical}
Next we analyze the stability of the states. We use the governing equations Eqs.~\eqref{x2dot}-\eqref{theta2dot}. We seek eigenvalues of the Jacobian matrix
\begin{equation}
M=
\begin{bmatrix}
Z_{x} & Z_{\theta}\\
Y_{x} & Y_{\theta}
\end{bmatrix},
\end{equation}
where $(Z_{x})_{ij} = \frac{\partial \dot{x_i}}{\partial x_j}$, $(Z_{\theta})_{ij} = \frac{\partial \dot{x_i}}{\partial \theta_j}$, $(Y_{x})_{ij} = \frac{\partial \dot{\theta_i}}{\partial x_j}$, and $(Y_{\theta})_{ij} = \frac{\partial \dot{\theta_i}}{\partial \theta_j}$ are evaluated from Eqs.~\eqref{x2dot}-\eqref{theta2dot}.

\subsubsection{Sync dots}
First we study the stability of the sync dots with a single cluster, i.e., the sync 1-dot. Table~\ref{table1} shows $S_{1+} = S_{1-} = S_{2+} = S_{2-} = S_{3+} =  S_{3-} = R_{1} = 1$. We only look at the order parameters that appear in Eqs.~\eqref{eq11}-\eqref{eq12}. The positions and phases are all synchronized and we get $x_i = \theta_i = c$, a constant that can be taken as $0$ without loss of generality. Putting all these into place, we calculate the eigenvalues of the Jacobian matrix $M$ at the fixed point $x_i = \theta_i = 0$. The eigenvalues are
\begin{equation}
	\lambda  = \begin{cases}
            0  &  \text{ multiplicity $2$} \\
		-J  &  \text{ multiplicity $N-1$} \\
		-K &  \text{ multiplicity $N-1$}
	\end{cases}.
\end{equation}
The sync 1-dot is stable for $J>0$ and $K>0$. Moving to the sync 2-dots state, here the fixed points look like $(x_i,\theta_i) = (c_1,c_2)$ for $i=1,2,\dots,N/2$, and $(x_i,\theta_i) = (c_1+\pi,c_2+\pi)$ for $i=N/2,\dots,N$. From Table~\ref{table1}, here $S_{1+} = S_{1-} = S_{3+} =  S_{3-} = 1$, and $S_{2+} =  S_{2-} = R_1 = 0$. The eigenvalues of the Jacobian $M$ at the fixed point yields
\begin{equation}
	\lambda  = \begin{cases}
            0  &  \text{ multiplicity $4$} \\
		-J/2  &  \text{ multiplicity $N-2$} \\
		-K/2 &  \text{ multiplicity $N-2$}
	\end{cases},
\end{equation}
when $N$ is even. If $N$ is odd then the eigenvalues are
\begin{equation}
	\lambda  = \begin{cases}
            0  &  \text{ multiplicity $4$} \\
		-\frac{(N+1)J}{2N}  &  \text{ multiplicity $\frac{N-1}{2}$} \\
            -\frac{(N-1)J}{2N}  &  \text{ multiplicity $\frac{N-1}{2}-1$} \\
		-\frac{(N+1)K}{2N}  &  \text{ multiplicity $\frac{N-1}{2}$} \\
            -\frac{(N-1)K}{2N}  &  \text{ multiplicity $\frac{N-1}{2}-1$} 
	\end{cases}.\label{2dot-odd}
\end{equation}
This shows sync 2-dots is also stable in the same region where the sync 1-dot is stable, i.e., $J>0,K>0$. 
\begin{figure}[hpt]
	\centering
	\includegraphics[width=\columnwidth]{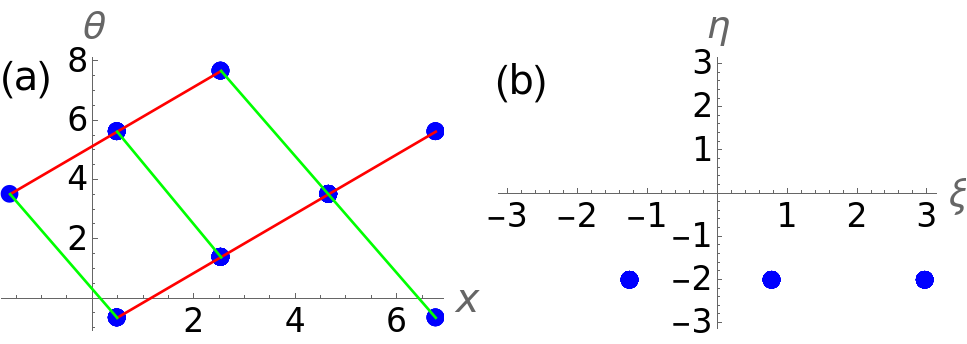}
	\caption{Visualization of the sync 3-dots. Simulation parameters: $(t,dt,N)=(200,0.1,200)$. (a) The positions and the phases are plotted against each other without restricting them to the interval $[-\pi,\pi]$. (b) They are plotted in the sum-difference coordinates after restricting them to  $[-\pi,\pi]$.}
	\label{pw3}
\end{figure}

The analysis becomes more intricate in the sync 3-dots state. We have already mentioned that the positions and phases are synchronized at three equidistant points having a difference $2\pi/3$ between them (Figs.~\ref{states-pulse}(c,f)). But if we look at their unrestricted positions and phases (meaning that they have not been pulled inside the interval $[-\pi,\pi]$), then their configuration looks like Fig.~\ref{pw3}(a) in the $x$-$\theta$ plane. The observable pattern is that the points can be joined in straight lines parallel to the lines $\theta=x$ and $\theta= -2x$. The ones parallel to $\theta=x$ are colored in red and the ones parallel to $\theta= -2x$ are colored in green. This is the reason that we get $S_{1-}=S_{2+}=1$ in this state~\footnote{Note that, in Table~\ref{table1}, $S_{1+}=S_{2+}=1$ as we assigned the maximum values to the `+' signs there. But, in reality, we get either $S_{1-}=S_{2+}=1$ or $S_{1+}=S_{2-}=1$} (we set rest of the order parameters to zero as they are much less than 1). The fixed points are of the form $(x_i, \theta_i) =  (c_1 + \frac{2\pi}{3} \ast i (\mod 3), \; c_2+ \frac{2\pi}{3} \ast i (\mod 3))$. Look at Fig.~\ref{pw3} for example. For finite $N$ the stability analysis yields tractable eigenvalues when $N$ is a multiple of 3, and we get following eigenvalues of $M$
\begin{equation}
	\lambda  = \begin{cases}
            0  \\
		\frac{1}{128} \left(-7 J-K \pm \sqrt{49 J^2-50 J K+K^2}\right)  \\
		\frac{1}{64} \left(-13 J-11 K \pm \sqrt{169 J^2-290 J K+121 K^2}\right)
	\end{cases},
\end{equation}
with multiplicities $2$, $4$ (including the $\pm$ signs), and $2N-6$, respectively. From this the stability of the sync 3-dots is calculated as $J>0$, $K>0$, which is same as the sync 1-dot and the sync 2-dots. The eigenvalues are calculated using a Mathematica notebook that we have provided with the references.

All $q>3$ dots do not exist for this sensing range ($p=2$) which completes our analysis of sync dots.

\subsubsection{Sync wave}
Here the phases are completely synchronized, but the positions are distributed along the ring. Order parameters takes values $R_1 = 1$, and $S_{n+} = S_{n-} = 0$ for $n=1,2,3$. The state is represented by the fixed point $(x_i,\theta_i) = (c_1+2\pi i/N, \; c_2)$. Setting $c_1$, $c_2$ to zero, the calculation of the eigenvalues of the Jacobian $M$ at this fixed point yields
\begin{equation}
	\lambda  = \begin{cases}
		0  &  \text{ multiplicity $a_1$} \\
		A(N) J   &  \text{ multiplicity $a_2$}\\
            -B(N) K   &  \text{ multiplicity $a_3$}
	\end{cases},
\end{equation}
where $A(N)$ and $B(N)$ are positive valued functions of $N$, and $a_1+a_2+a_3=2N$. This reveals the sync wave is stable if $J<0$ and $K>0$.

\subsubsection{$1$-wave}
1-wave corresponds to $\theta = \pm x +C$, for some constant $C$ (determined by the initial conditions). If we choose the `$+$'ve sign, then order parameters become $S_{1-} = 1$ and $S_{1+} = S_{2+} = S_{2-} = S_{3+} = S_{3-} = R_1 = 0$ (choice of the `$-$' sign just alters the values of $S_{1+}$ and $S_{1-}$ while leaving the analysis unaffected). The fixed point for this state is $(x_i,\theta_i) = (c_1 + 2\pi i/N, \; c_2+ 2\pi i/N )$. Jacobian gives the eigenvalues
\begin{equation}
	\lambda  = \begin{cases}
		0  \\
		\frac{1}{32} (-5 J-7 K) \\
            \frac{1}{64} \left(-J-K \pm \sqrt{J^2+34 J K+K^2}\right)\\
            \frac{1}{64} \left(-J-3 K \pm \sqrt{J^2+246 J K+9 K^2}\right)\\
            \frac{1}{64} \left(-J-5 K \pm \sqrt{J^2+298 J K+25 K^2}\right)\\
            \frac{1}{128} \left(-7 J-13 K\pm \sqrt{49 J^2+566 J K+169 K^2}\right)
	\end{cases},
\end{equation}
with multiplicities $N-7$, $N-9$, $4$, $4$, $4$, and $4$, respectively that add up to $2N$, the dimension of the whole system. From here, we get the 1-wave is stable in the region where $J+K>0$ and $J+5K>0$, and one of $J$ and $K$ is less tan zero. The eigenvalues are calculated assuming that $N$ is sufficiently large ($N \geq 10$ in this case). It should be noted that the part of stability region of the 1-wave that lies in the second quadrant of the $J$-$K$ plane, falls inside the stability region of the sync wave. This is the reason that 1-wave is bistable with the sync wave inside this region (see Fig.~\ref{jk-space}). We discuss the bistability in details in Sec.~\ref{bistable}.

\subsubsection{$2$-wave}
Here, we have $S_{2+} = 1$ and $S_{1+} = S_{1-} = S_{2-} = S_{3+} = S_{3-} = R_1 = 0$. This corresponds to the fixed points $(x_i,\theta_i) = (c_1+2\pi i/N, \; c_2+2 \ast 2\pi i/N)$. We calculate the eigenvalues of $M$
\begin{equation}
	\lambda  = \begin{cases}
		0  \\
		\frac{1}{8} (-2 J-K) \\
            \frac{1}{8} \left(-J \pm \sqrt{J (J+2 K)}\right)\\
            \frac{1}{32} \left(-2 J \pm \sqrt{J (4 J+3 K)}\right)\\
            \frac{1}{32} \left(-2 J-K \pm \sqrt{4 J^2+68 J K+K^2}\right)\\
            \frac{1}{128} \left(-11 J-K \pm \sqrt{121 J^2+1174 J K+K^2}\right)\\
            \frac{1}{128} \left(-13 J-7 K \pm \sqrt{169 J^2+502 J K+49 K^2}\right)
	\end{cases},
\end{equation}
with multiplicities $N-9$, $N-11$, $4$, $4$, $4$, $4$, and $4$, respectively. Then we calculate the region where the real part of the eigenvalues are negative and we get the condition of stability as $13J+7K>0$ and $K<0$. Note that, this region contains the part of the stability region of the 1-wave that lie in the fourth quadrant of the $J$-$K$ plane, and this is why we encounter bistability between the 1-wave and 2-wave here along with occasional occurrence of the active state. Look at Fig.~\ref{jk-space} for details.

\subsubsection{$3$-wave}
Solutions are now of the form $\theta = \pm 3x +C$. Taking the `$-$'ve sign (without loss of generality), we get $S_{3+} = 1$ and $S_{1+} = S_{1-} = S_{2+} = S_{2-} = S_{3-} =  R_1 = 0$. Fixed point is of the form $(x_i,\theta_i) = (c_1+2\pi i/N, \; c_2+3 \ast 2\pi i/N)$. We were unable to perform the stability analysis in this case as the expressions became humongous. Numerics revealed that the 3-wave shares the boundary with the async state. Next, we analyze the stability of the async state.

\subsubsection{Async state}
We study the async state in the thermodynamic limit, $N \to \infty$.
The density function satisfies the continuity equation
\begin{equation}
    \frac{\partial \rho}{\partial t} + \nabla \cdot (v\rho) = 0, 
\end{equation}
where $\rho(x,\theta,t)$ denotes the probability that a swarmalator is at position $x$ with its phase $\theta$ at time $t$, and the velocity $v=(v_x,v_{\theta})$ is the right hand side of Eqs.~\eqref{eq11}-\eqref{eq12}.
Consider a perturbation around the static async state $\rho_0 = 1/(4\pi^2)$,
\begin{equation}
	\rho = \rho_0 + \epsilon \rho_1 = 1/4\pi^2 + \epsilon \rho_1.
\end{equation}
The velocity $v_0=0$ in the async state. The perturbation of velocity thus become $v = v_0 + \epsilon v_1$ = $\epsilon v_1$. The perturbed density and velocity follow the following equation
\begin{equation}
	\frac{\rho_1}{\partial t} + \rho_0 \nabla \cdot v_1 = 0,
\end{equation}
and the perturbed order parameters of interest become
\begin{align}
	W_{n \pm}^1 &=  \int e^{i (n x' \pm \theta' )} \rho_1(x',\theta') dx' d\theta', \; \; n=1,2,3, \nonumber \\
	Z_1^1 & = \int e^{i \theta'} \rho_1(x',\theta') dx' d\theta' \label{pert}.
\end{align}
After carrying out the analysis and comparing the Fourier modes, we achieve
\begin{align}
    \Dot{W}_{1{\pm}} &\propto (5J+7K) W_{1{\pm}}, \nonumber\\
    \Dot{W}_{2{\pm}} &\propto (2J+K) W_{2{\pm}}, \nonumber\\
    \Dot{W}_{3{\pm}} &\propto (3J+K) W_{3{\pm}}, \nonumber\\
    \Dot{Z}_1 &\propto K Z_1.
\end{align}
From these, we derive that the async state is stable when $3J+K<0$ and $K<0$.

\begin{figure}[t]
	\centering
	\includegraphics[width=\columnwidth]{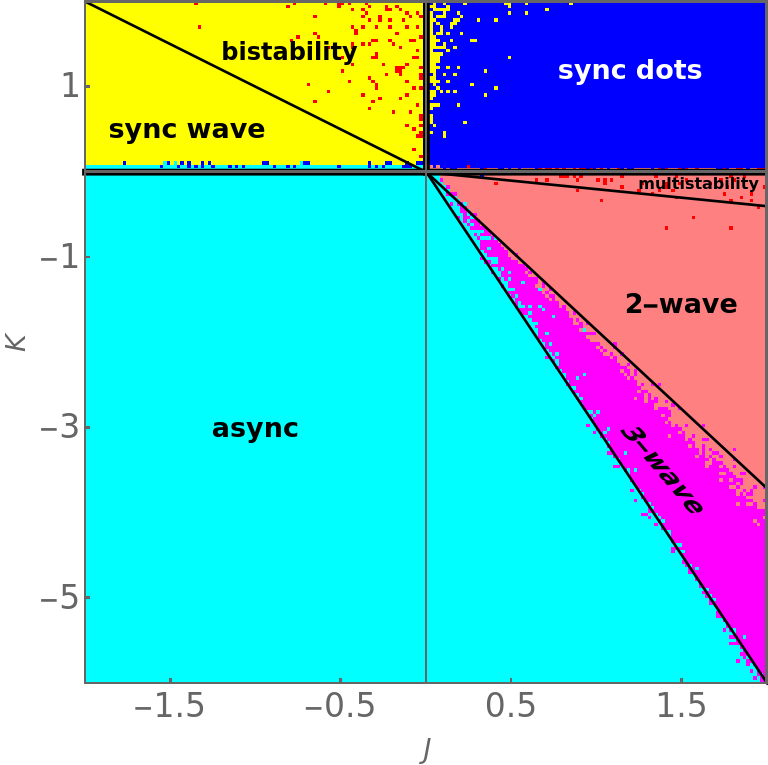}
	\caption{Bifurcation diagram in the $J$-$K$ plane. At $t=0$, positions and phases are chosen randomly from $[0,2\pi]$. The region is divided into $100$ by $100$ grid points and at each point simulation is performed with $(t,dt,N) = (500,0.1,200)$. All the order parameters are averaged over last $10\%$ data. The black lines are analytical predictions.}
	\label{jk-space}
\end{figure}

With the wisdom of the order parameters defined in Table~\ref{table1} and knowledge about the stability, we proceed to study the numerical bifurcation scenario in the $J$-$K$ plane and look to validate our analytical findings. We choose the ranges of $J$ and $K$ from intervals where all the collective states occur. $J\in [-2,2]$ and $K \in [-6,2]$ serve our purpose. The sync dots are found when $J>0$ and $K>0$, the blue region of Fig.~\ref{jk-space}. Sync wave prevails when $K>0$ and $J<0$ (yellow region). We observe the occurrence of 2-wave in the pink region where $-2J<K<0$. From our analysis, we expected to find 1-wave in the region $5J+7K>0$, the area above the black dotted line. But, we see that 1-wave is bistable with the 2-wave in the fourth quadrant, with the sync wave in the second quadrant. The probability of 1-wave occurring is much less than the 2-wave and the sync wave state. The scattered red dots correspond to points in the parameter space where 1-wave is found. 3-wave is found to exist in the anticipated region $-3J<K<2J$ (magenta region). The async state occupies a vast area of the parameter space where $K<-3J$ and $K<0$ (cyan region) and it validates our analysis as well.

\subsection{Bistability} \label{bistable}
\begin{figure}[hpt]
	\centering
	\includegraphics[width=\linewidth]{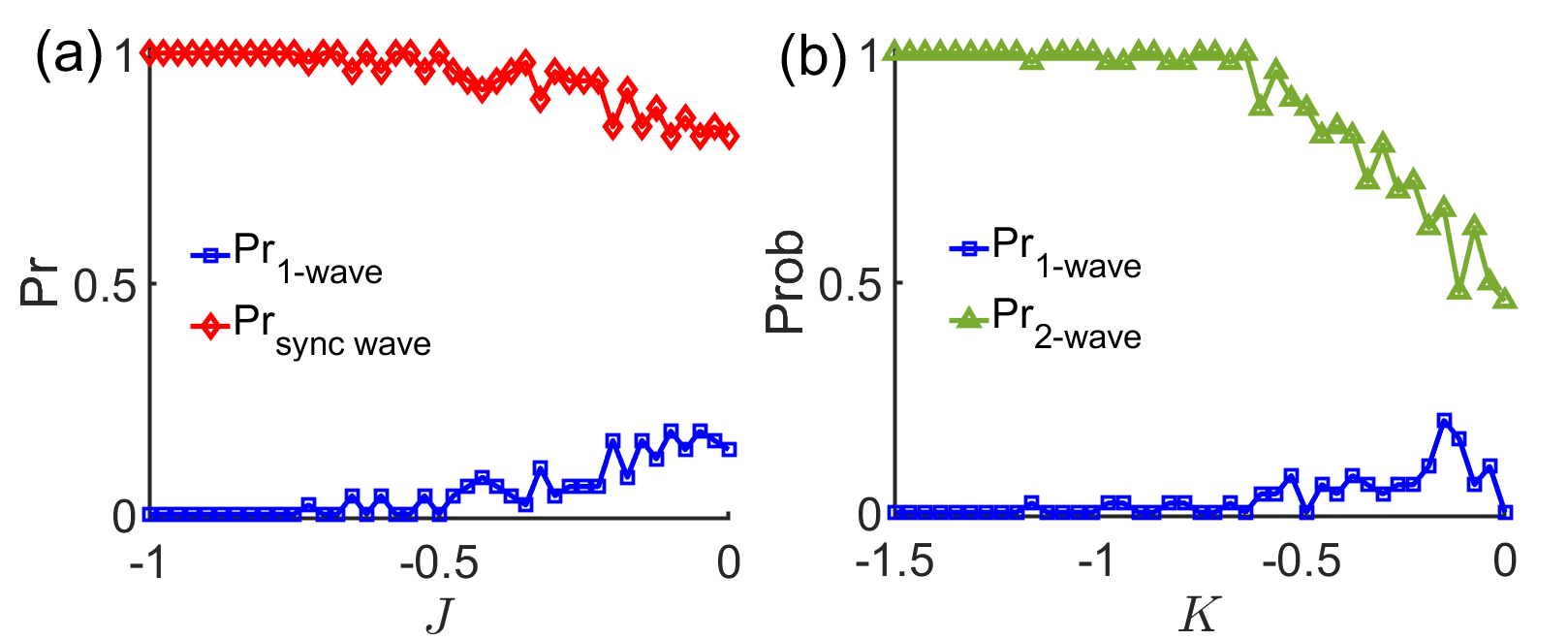}
	\caption{Bistability of the 1-wave with the sync wave and 2-wave. (a) Here, $K=1.0$ and we vary $J$ where 1-wave and sync wave are bistable. (b) Here, $J=2.5$ and $K$ is varied to check the bistability between the 1-wave and 2-wave. For simulations, we use $(t,dt,N)=(200,0.1,200)$. The probabilities are calculated by taking $50$ different initial conditions.}
	\label{bistabilty}
\end{figure}

We found in the last section that the 1-wave is bistable with the sync wave and the 2-wave. To scrutinize this, we calculate the probability of finding the 1-wave along with the sync wave and 2-wave. We run the simulation $q$ times and count how many times the system reaches these states and the probabilities are calculated accordingly. Figure~\ref{bistabilty} showcases the results. In Fig.~\ref{bistabilty}(a), $K$ is fixed at $1.0$ and $J$ is varied from $-1$ to $0$. 1-wave (blue squares) is found to be bistable with the sync wave (red diamonds) here. In Fig.~\ref{bistabilty}(b), $J=2.5$ and $K$ is varied from $-1.5$ to $0$. Now, 1-wave is bistable with 2-wave (green triangles). In both the cases, the probability of 1-wave is much lesser than the other one. The probability is almost zero at the beginning and only increases as $J$ or $K$ tends to zero. The other noticeable thing from Fig.~\ref{bistabilty}(b) is that, around $K=0$, the probabilities do not add up to 1. This indicates the existence of a third state different from the 1-wave and 2-wave. This is actually the active state that occurs near the transition boundary which we have mentioned earlier. 

\section{Variable length scale $p$}

Let's distill what we found so far. We froze the coupling range at $p=2$ and varied the parameters $J,K$. This revealed several new states: $q-$waves, sync dots, sync wave and active state. Figure~\ref{jk-space} shows us where these occur in the $(J,K)$ plane, and also highlights the multi-stability regions.

What happens when coupling range $p$ becomes the active parameter? In particular, what happens when $p >> 1$ which takes us into the short-range regime? 

Somewhat surprisingly, we find that no new states occur. All that changes is the maximum winding number $q_{max}$ for both the sync dots and $q$-waves increases. Its easy to see why. You can see that the largest harmonic $n_{max}$ in the coupling kernel $G(x) = (1+\cos x)^p/2^p$ scales linearly with the range $n_{max} = p$. In turn, this largest harmonic determines the $q_{max}$ of both the $q$-waves and sync dots. Take the $q=5$-wave. This requires the $5$-th rainbow order parameter to be unity $S_{5}=1$ while all the others are zero. And in order for the $S_{5}$ to exist in the EOM's, we need at least $p>=4$ to generate the required terms. The same argument holds for the $q>1$ sync dots.

Now we calculate how the states' stability thresholds depend on the coupling range $p$. The spirit of the analysis is the same as before. All that changes is there are more trigonometric harmonics. The space and phase equations, for example, become
\begin{multline}
	\dot{x_i} = \frac{J}{N} \sum_{j=1}^{N} \sum_{m=1}^{p+1} A_{m-} \Bigg[\sin \Big( m(x_j-x_i) + (\theta_j-\theta_i) \Big) \\ + \sin \Big( m(x_j-x_i) - (\theta_j-\theta_i) \Big) \Bigg], \label{newx}
\end{multline}
\begin{multline}
	\dot{\theta_i} = \frac{K}{N} \sum_{j=1}^{N} \Bigg[ B \sin \Big( \theta_j - \theta_i\Big) \\ + \sum_{m=1}^{p+1} A_{m+}  \Big[\sin \Big( m(x_j-x_i) + (\theta_j-\theta_i) \Big)  \\ - \sin \Big( m(x_j-x_i) - (\theta_j-\theta_i) \Big)\Big] \Bigg] , \label{newtheta}
\end{multline}
where
\begin{equation}
	A_{m \pm}  = \begin{cases}
		\frac{1}{2^{2p+1}} \Big[\binom{2p}{p+1-m} \pm \binom{2p}{p-1-m}\Big], &  1 \le m \le p-1\\
		\frac{2p}{2^{2p+1}}, & m =p \\
		\frac{1}{2^{2p+1}}, &m=p+1
	\end{cases},
\end{equation}
and
\begin{equation}
	B = \frac{1}{2^{2p}} \binom{2p}{p-1}.
\end{equation}
Here we have used the identity
\begin{equation}
	\left( \frac{1 + \cos y}{2}\right)^r = \frac{1}{2^{2r}} \binom{2r}{r} + \frac{1}{2^{2r-1}} \sum_{k=0}^{r-1} \binom{2r}{k} \cos ((r-k)y).
\end{equation}
which maps powers of $\sin^m x$ to $\sin(m x)$ (the same for cosines). It is elementary to show that the coefficients $A_{m\pm}, B>0$.

{\bf Stability of the async:} We will have 
\begin{align}
	\Dot{W}_{q{\pm}} &\propto (q A_{q-} J + A_{q+}K) W_{q{\pm}},
\end{align}
for $q=1,2,\dots, p+1$, and 
\begin{align}
	\Dot{Z}_1 &\propto K Z_1.
\end{align}
After carrying out the calculations, we find the async state's stability as
\begin{align}
(p+1)J+K<0, \\
 K<0.
\end{align}

{\bf Stability of the sync dots: } Fixed points is $x_i = c_1, \theta_i=c_2$ when swarmalators are synchronized to one dot. Eigenvalues are 
\begin{equation}
	\lambda  = \begin{cases}
		0  &  \text{ multiplicity $2$} \\
		-J &  \text{ multiplicity $N-1$} \\
            -K &  \text{ multiplicity $N-1$}
	\end{cases}.
\end{equation}
Hence, stable when $J,K>0$. In the sync 2-dots, positions and phases are synchronized to two points at $\pi$ difference. The eigenvalues are
\begin{equation}
	\lambda  = \begin{cases}
		0  &  \text{ multiplicity $4$} \\
		-J/2 &  \text{ multiplicity $N-2$} \\
            -K/2 &  \text{ multiplicity $N-2$}
	\end{cases},
\end{equation}
assuming $N$ is even. When $N$ is odd the eigenvalues are same as Eq.~\eqref{2dot-odd}. This reveals that the sync 2-dots is also stable when $J,K>0$. The eigenvalues of the sync 3-dots are comparatively difficult to derive, but when checked numerically, are also found to be negative in the real part in the region $J,K>0$.

{\bf Stability of the sync wave:} Fixed point is $x_i = 2 \pi i/N + c_1, \theta_i = c_2$. Eigenvalues are
\begin{equation}
	\lambda  = \begin{cases}
		0  &  \text{ multiplicity $a_1$} \\
		A(N,p) J   &  \text{ multiplicity $a_2$}\\
            -B(N,p) K   &  \text{ multiplicity $a_3$}
	\end{cases},
\end{equation}
where $A(N,p)$ and $B(N,p)$ are positive valued functions of $N$ and $p$, and $a_1,a_2,a_3$ are the multiplicities such that $a_1+a_2+a_3=2N$. So we infer, sync wave is stable when $J<0$ and $K>0$.

{\bf Stability of the $q$-waves:}. Unfortunately, we were unable to find the stability here for general $p$. Numerics however indicate, like the $p=2$ case, they bifurcate from the sync dots and async states, and so in that sense we have their stability regions \footnote{In theory they could be bistable with the sync dots or async state of course, but numerics indicate this is not the case}.

\begin{figure}[hpt]
	\centering
	\includegraphics[width=\linewidth]{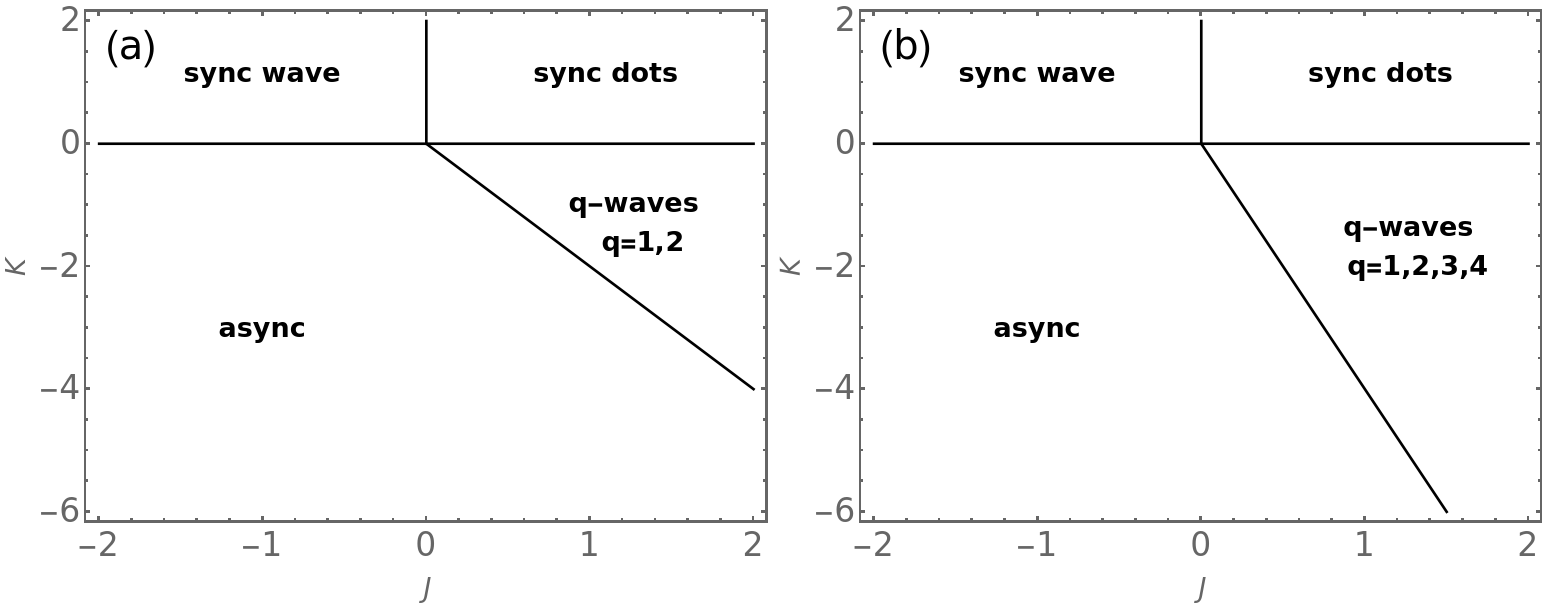}
	\caption{Locations in the $(J,K)$ plane where the steady states are stable (a) $p=1$, (b) $p=3$. Boundaries are calculated analytically. See text for details.}
	\label{bif-diagram-local-c}
\end{figure}

\begin{figure}[hpt]
	\centering
	\includegraphics[width=\linewidth]{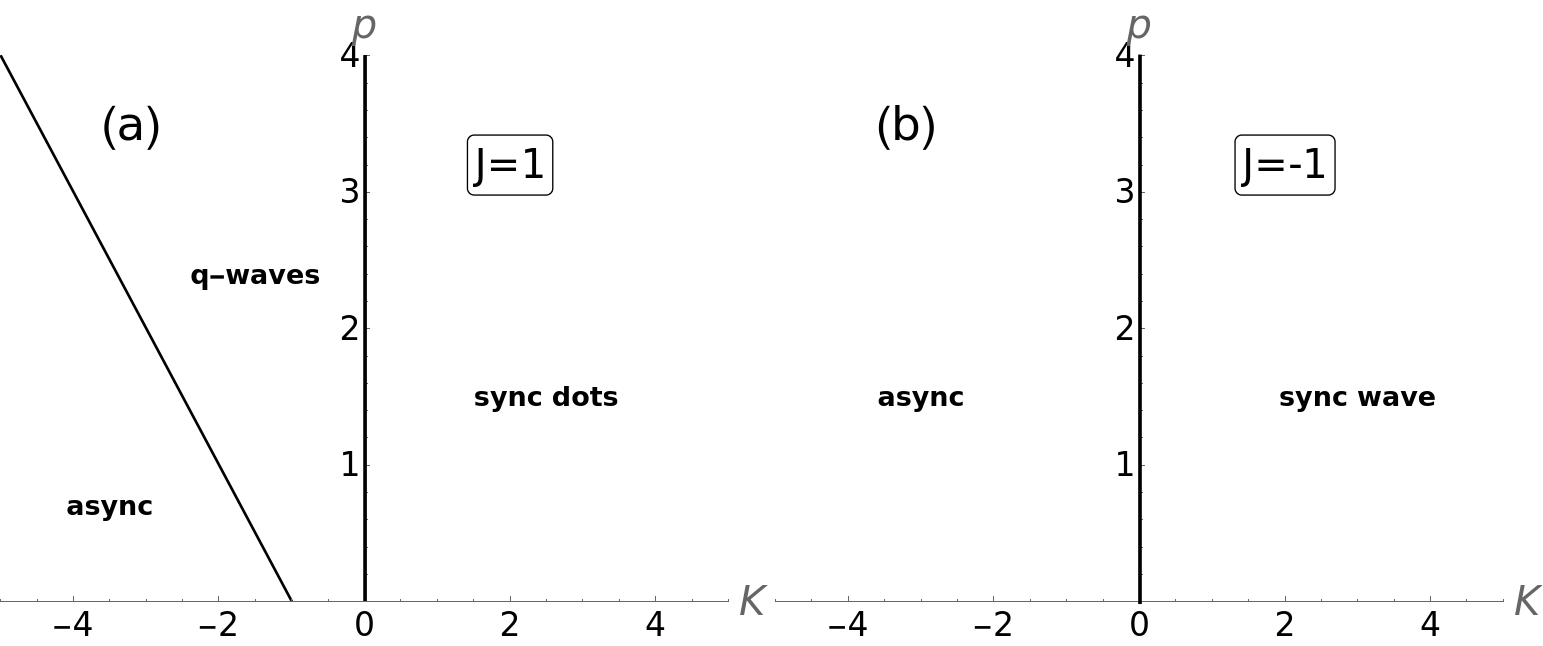}
	\caption{Locations in the $(K,p)$ plane where the steady states are stable. (a) $J=1$. (b) $J=-1$. Boundaries are calculated analytically. See text for details.}
	\label{bif-diagram-local-d}
\end{figure}

We summarize our results in two ways. First, we show bifurcations diagrams in the $(J,K)$ plane for $p=1,3$ (Fig.~\ref{bif-diagram-local-c}) which can be compared to the $p=2$ case. You can see the effect or larger $p$, which corresponds to lower sensing range, is that new $q$-waves are born that `cut' into the stability region of the async state.  
Second, we show the bifurcation diagram in the $(K,p)$ plane for both $J=1,-1$ (Fig.~\ref{bif-diagram-local-d}) which displays the same information.

\section{Discussion}
Our contribution is the first analytic study of swarmalators with a controllable coupling range. Our primary finding was that lowering the range led to several new collective states, such as sync dots, sync wave, $q$-waves, and an active state. We drew the bifurcation diagram for the system which revealed regions of multistability. Table~\ref{table2} catalogs the states we found along with the states of the baseline model (recovered when $p=0$).

We suspect some of these states might be observable in real world swarmalator systems which swarm in 1d, such as bordertaxic vinegar eels or sperm. This systems sometimes display metachronal waves equivalent to our $q$-waves \cite{quillen2021metachronal,quillen2022fluid}. The states may also be realizable in synthetic swarmalator systems, such as active spheres \cite{riedl2024entropy} or colloids \cite{ketzetzi2022activity}. 

Future work could extend our results from the 1d model to the original 2d swarmalator model, which is defined on the open plane \cite{o2017oscillators}, or a simpler 2d model which is defined on the plane with periodic boundary conditions   \cite{o2023solvable}. Adding heterogeneity to the natural frequencies or coupling constants would also be interesting.

\begin{table}[t!]
\centering
\renewcommand{\arraystretch}{1.5}
\setlength{\tabcolsep}{12pt}
\begin{tabular}{|c|c|}
\hline
\textbf{Global range ($p=0$)} & \textbf{Non-global range ($p>0$)} \\ \hline 
sync dots ($k=1,2$)                    & sync dots ($k=1,2$)                  \\ \hline
$\times$                    & sync dots ($k>=3$)                  \\ \hline
async                         & async                       \\ \hline
phase wave ($k=1$)            & phase wave ($k=1$)          \\ \hline
               $\times$               & phase waves ($k>1$)         \\ \hline
                  $\times$            & sync wave                   \\ \hline
\end{tabular}
\caption{\label{table2}Comparison between the states for the global and local range behavior of the coupling kernel $G(x)$.}
\end{table}


\section*{Code availability}
Codes used for numerical simulations in this paper are openly available at the GitHub repository~\footnote{https://github.com/Khev/swarmalators/tree/master/1D/on-ring/local-coupling}.

\bibliographystyle{apsrev}

\end{document}